\documentclass[twocolumn,superscriptaddress,amsmath,amssymb,aps,floatfix,longbibliography]{revtex4}

\usepackage{graphicx}
\usepackage{tabularx}
\usepackage{siunitx}
\usepackage[dvipsnames]{xcolor}

\usepackage{xcolor}

\begin{document}

\newcommand{\mum}[1]{${#1}$\,\textmu m} 
\newcommand{\ED}[1]{\textcolor{magenta}{\,#1}}
\newcommand{\RS}[1]{\textcolor{red}{\,#1}}
\newcommand{\DG}[1]{\textcolor{orange}{\,#1}}
\newcommand{\SY}[1]{\textcolor{blue}{\,#1}}
\newcommand{\LDL}[1]{\textcolor{orange}{\,#1}}
\newcommand{\YX}[1]{\textcolor{ForestGreen}{\,#1}}

\title{Dehydration drives damage in the freezing of brittle hydrogels}

\author{Shaohua Yang}
\thanks{These two authors contributed equally}
\affiliation{School of Mechanical Engineering and Automation, Beihang University, Beijing, China.}%
\affiliation{Department of Materials, ETH Z\"{u}rich, 8093 Z\"{u}rich, Switzerland.}

\author{Dominic Gerber}
\thanks{These two authors contributed equally}
\affiliation{Department of Materials, ETH Z\"{u}rich, 8093 Z\"{u}rich, Switzerland.}%

\author{Yanxia Feng}
\affiliation{Department of Materials, ETH Z\"{u}rich, 8093 Z\"{u}rich, Switzerland.}%

\author{Nicolas Bain}
\affiliation{University of Lyon, Universit\'{e} Claude Bernard Lyon 1, CNRS, Institut Lumi\`{e}re Mati\`{e}re, F-69622, Villeurbanne, France.}%

\author{Matthias Kuster}
\affiliation{Department of Mechanical and Process Engineering, ETH Z\"{u}rich, 8093 Z\"{u}rich, Switzerland.}%

\author{Laura de Lorenzis}
\affiliation{Department of Mechanical and Process Engineering, ETH Z\"{u}rich, 8093 Z\"{u}rich, Switzerland.}%

\author{Ye Xu}
\email{ye.xu@buaa.edu.cn}
\affiliation{School of Mechanical Engineering and Automation, Beihang University, Beijing, China.}%

\author{Eric R. Dufresne}
\email{eric.r.dufresne@cornell.edu}
\affiliation{Department of Materials Science \& Engineering, Cornell University, Ithaca, USA}%
\affiliation{Laboratory of Atomic and Solid State Physics, Cornell University, Ithaca, USA}

\author{Robert W. Style}
\email{robert.style@mat.ethz.ch}
\affiliation{Department of Materials, ETH Z\"{u}rich, 8093 Z\"{u}rich, Switzerland.}%

\begin{abstract}
It is widely known that freezing breaks soft, wet materials.
However, the mechanism underlying this damage is still not clear.
To understand this process, we freeze model, brittle hydrogel samples, while observing the growth of ice-filled cracks that break these apart.
We show that damage is not caused by the expansion of water upon freezing, or the growth of ice-filled cavities in the hydrogel.
Instead, local ice growth dehydrates the surrounding hydrogel, leading to drying-induced fracture.
This dehydration is driven by the process of cryosuction, whereby undercooled ice sucks nearby water towards itself, feeding its growth.
Our results highlight the strong analogy between freezing damage and desiccation cracking, which we anticipate being useful for developing an understanding of both topics.
Our results should also give useful insights into a wide range of freezing processes, including cryopreservation, food science and frost heave.
\end{abstract}

\maketitle

Damage caused by the freezing of soft, wet materials is a widely important process.
It hinders our ability to cryopreserve tissue \cite{tas_freezer_2021}, and has implications for plant and animal life at cold temperatures \cite{benson_cryopreservation_2008,knight2001adsorption}.
It is a key consideration in the cold storage of food and medicine \cite{petzold_ice_2009,deville_freezing_2017,matthias2007freezing}, causes expensive deterioration to infrastructure in cold climates \cite{peppin_physics_2013,kjelstrup_transport_2021}, and shapes periglacial landscapes \cite{dash_physics_2006}.
Despite its widespread importance, and its long history of research, freezing damage is still poorly understood.
For example, when freezing a particular soil type, it is not possible to predict how fast, and in what form ice will grow \cite{dash_physics_2006,peppin_physics_2013,style_kinetics_2012} -- despite the availability of a wide range of frost-heave models \cite{konrad_mechanistic_1980,kjelstrup_transport_2021,wettlaufer_theory_1996,you_situ_2018,zhou_ice_2020,rempel_microscopic_2011}.
Thus, our knowledge of freezing damage is largely empirical.

There are several key problems that make understanding the freezing process challenging.
Freezing is affected by multiple factors, including temperature gradient, solute concentration, type of solute, and material porosity \cite{gerber_stress_2022,schollick_segregated_2016,dedovets_fivedimensional_2018}, and this makes it hard to isolate the key physics.
Furthermore, most experiments are done on bulk, opaque samples (e.g. \cite{taber_mechanics_1930,konrad_mechanistic_1980}), and analyzed via approaches such as observing macroscopic sample changes \cite{arenson2008new}, sectioning frozen samples \cite{hagiwara_fractal_2002}, tracking ice growth with techniques like differential scanning calorimetry \cite{devireddy1998measurement} or magnetic resonance imaging \cite{watanabe_amount_2002}.
Although these methods give useful insights, they cannot visualise the crucial microscopic ice-growth processes that underlie damage.
Finally, commonly frozen materials like clays, colloidal suspensions or tissue are often heterogeneous or have complex rheological properties, further obscuring the picture \cite{peppin_morphological_2007,deville_influence_2010}.

To overcome these issues, we perform freezing experiments on model, transparent hydrogels.
We observe samples with a confocal microscope \cite{dedovets_fivedimensional_2018,gerber_stress_2022,gerber2023polycrystalinity}, allowing us to directly visualise 3-D ice-crystal growth and the resulting hydrogel damage.
Surprisingly, this damage is caused by growing ice crystals dehydrating the surrounding hydrogel, leading to desiccation cracking.

\begin{figure}[htbp]
    \centering
    \includegraphics[width=1\linewidth]{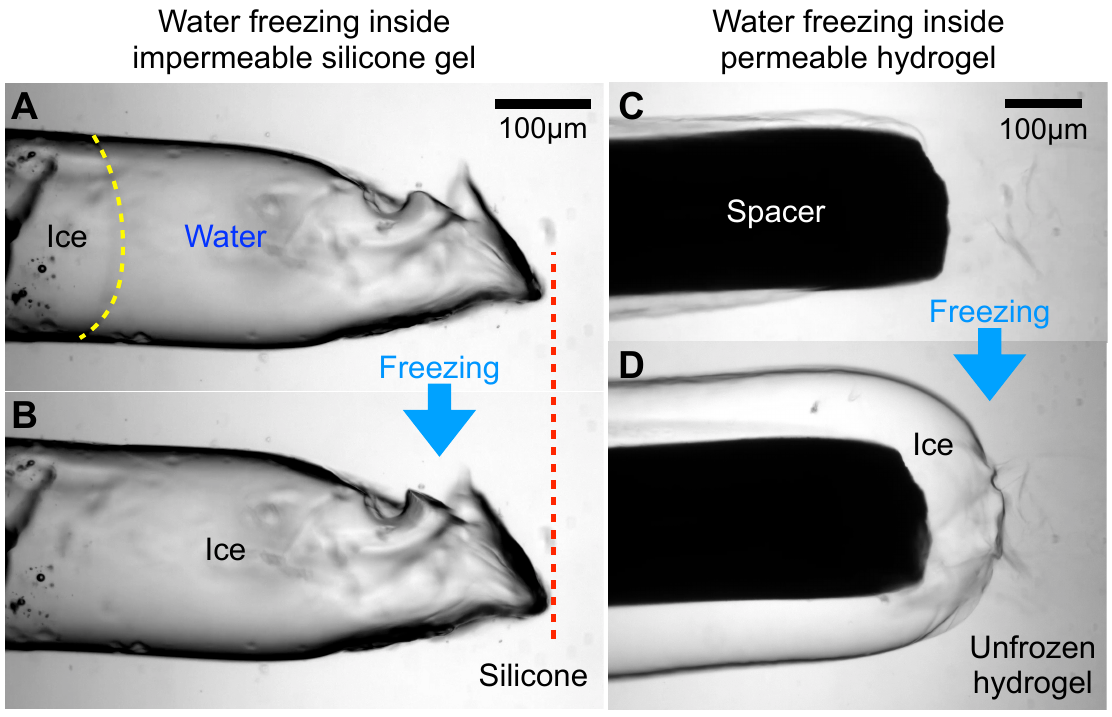}
    \caption{Freezing damage in soft materials is not caused by the expansion of water upon freezing. (A,B) Freezing of water in a cavity in a silicone rubber ($\mu=30$ kPa) causes negligible deformation to the rubber. The deformation is caused by the expansion of water upon freezing. (C,D) Freezing of a brittle, PEGDA hydrogel (90\% water, $\mu=37$ kPa) containing a metal spacer. Ice forms around the spacer, and continues to grow by sucking in water from the surrounding hydrogel \emph{via} cryosuction, leading to large stresses in the gel.}
    \label{fig:schem}
\end{figure}

\begin{figure*}[htbp]
    \centering
    \includegraphics[width=\linewidth]{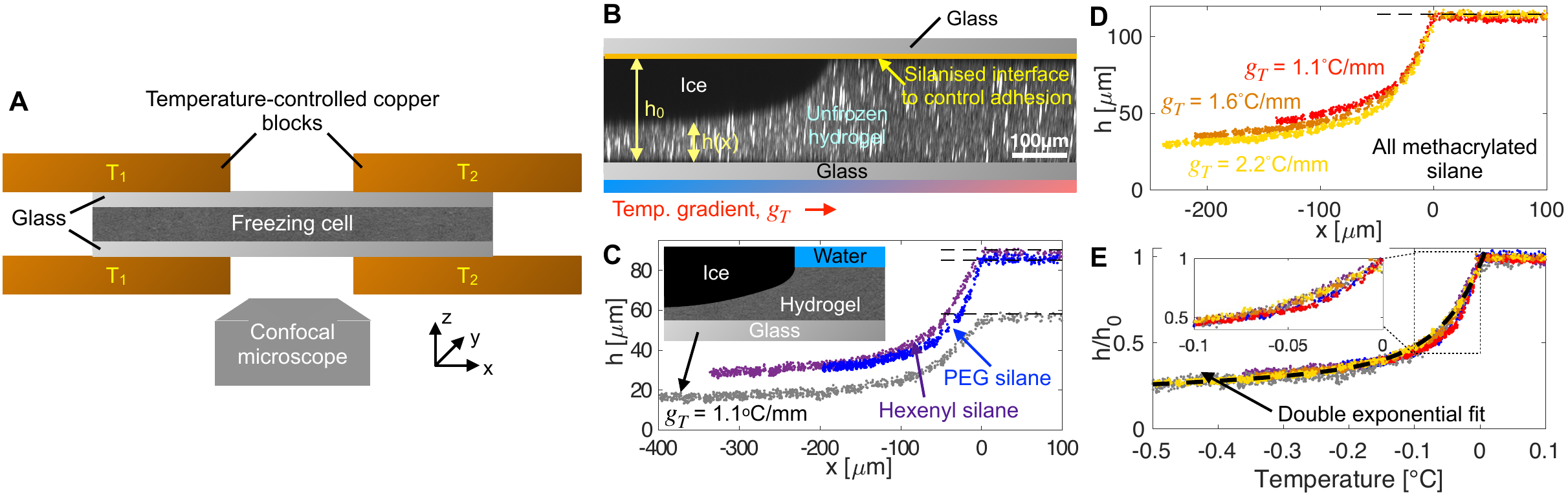}
    \caption{Freezing of PEGDA hydrogels in a temperature gradient. (A) Schematic of the freezing apparatus. (B) A side view of a typical experiment, imaged with confocal microscopy. Bright spots are fluorescent tracer particles embedded in the hydrogel. Upon freezing, ice breaks the silanized, top interface between the hydrogel and the glass cell.
    (C) Hydrogel delamination profiles for three different interfacial strengths: strong, hexenyl silane (purple), weak, PEG silane (blue) and no interface (grey). The schematic shows the geometry for the last case. 
    (D) Hydrogel delamination profiles for a strong interface (methacrylated silane) at three different temperature gradients. In all experiments in (C,D), the hydrogel shape reaches a steady state after $\sim 20$ mins, so we image samples after 40 mins.
    (E) The data from (C,D) largely collapse when local thickness is scaled with initial hydrogel thickness, and plotted against temperature. The dashed curve shows an empirical fit, described in the main text. The inset zooms in on the data near $T=0$, showing small deviations from a perfect collapse.}
    \label{fig:profiles}
\end{figure*}

It is important to emphasize that damage in soft materials is not typically caused by the $\sim9\%$ volumetric expansion of water upon freezing.
We demonstrate this by observing ice growth in a cavity in a water-impermeable silicone gel (Figure \ref{fig:schem}A,B).
As water freezes in the cavity, there is a small expansion (see dashed line).
However, this expansion --  2\% in each dimension based on water's volumetric expansion upon freezing -- is easily accommodated by the gel.
Most soft materials can easily tolerate such strains.

Instead, damage is driven by \emph{cryosuction}, whereby water is sucked towards ice at subzero temperatures, feeding ice growth \cite{gerber_stress_2022}.
In Figure \ref{fig:schem}C,D, we compare ice growth in a poly(ethylene glycol) diacrylate (PEGDA) hydrogel of comparable shear modulus, $\mu$, to the silicone gel.
Now ice growth in the gel is extensive, eventually ripping apart the hydrogel (Supplementary Video 1).
This growth is driven by cryosuction of water from the unfrozen hydrogel surrounding the ice.
The water in the hydrogel does not freeze, as capillarity (the Gibbs-Thomson effect) prevents ice from growing into the hydrogel mesh until well below the bulk freezing temperature \cite{dash_physics_2006}.

To understand how such freezing damage occurs, we freeze hydrogel-filled cells in a linear temperature gradient, $g_T$ (Figure \ref{fig:profiles}A), and examine the resulting fractures.
We use a freezing stage on a confocal microscope (Figure \ref{fig:profiles}A \cite{gerber_stress_2022,gerber2023polycrystalinity}) allowing us to control temperature profiles.
Here, we impose either isothermal conditions or fixed temperature gradients along the $x-$direction.
In a fixed, small temperature gradient, ice grows into the hydrogel from the cold side of the cell, fracturing apart the top hydrogel/glass interface.
A side-view of a typical delamination, obtained via confocal microscopy, is shown in Figure \ref{fig:profiles}B -- bright spots are fluorescent nanoparticles embedded in the hydrogel.
This crack shape is essentially 2-D, varying little in the $y-$direction, and reaches a steady-state within 20 minutes (see Supplementary Material).
Thus, such cracks are very convenient for studying freezing damage.

The overall shape of the delamination in Figure \ref{fig:profiles}B is reminiscent of a classical crack -- with the large delamination opening suggesting that ice is aggressively pushing apart the glass/hydrogel interface.
However, as we will see, this is not correct:  the large delamination opening away from the crack tip is actually mainly caused by temperature-dependent hydrogel shrinkage. The shape is largely independent of fracture parameters such as adhesion strength, and thus
does not immediately tell us about how the ice drives fracture. 
We demonstrate this by varying parameters such as how strongly adhered the hydrogel/glass interface is, and the temperature gradient applied to the sample.
We change hydrogel/glass adhesion using different silane pre-treatments of the glass.
PEG silane (3-[methoxy(polyethyleneoxy)9-12]propyltrimethoxysilane) is an unreactive, hydrophilic coating that should result in very low adhesion strength.
Hexenyl silane (5-hexenyltrimethoxysilane) is a coupling agent containing a vinyl group that should chemically bind to the hydrogel, yielding a much higher adhesion strength.
We compare crack shapes for the different silanes (blue and purple data in Figure \ref{fig:profiles}C, both with $g_T=1.1^\circ$C/mm).
Here, dashed lines show the original sample thicknesses, and the crack tip is located at $x=0$.
Away from the crack tip, these shapes are very similar, suggesting that they are independent of adhesion strength.
Only when we look closer to the crack tip do we see evidence of sensitivity to adhesion.
By contrast, the shape away from the crack tip does depend on the temperature gradient in the sample.
Figure \ref{fig:profiles}D shows crack shapes for $g_T=1.1,\, 1.6,\, 2.2 ^\circ$C/mm.
These use a methacrylated silane (3-(trimethoxysilyl)propyl methacrylate, or `bind-silane'), which is another coupling agent that should give good glass/hydrogel adhesion.
As we increase $g_T$, the overall crack shapes progressively open -- except within about 50$\mu$m of the crack tip, where the shape appears to be fixed.

The large-scale crack shape is predominantly dependent on local temperature.
In Figure \ref{fig:profiles}E, we replot the data from Figures \ref{fig:profiles}C,D by scaling $h(x)$ by initial thickness, $h_0$ and plotting this against the temperature, $T=g_Tx$.
There is an excellent collapse of all the data, excepting some small deviations near the crack tip (see inset).
Indeed, we see the same, temperature-dependent shape even when there is no crack at all:
when we freeze a hydrogel layer that is simply submerged in water (Figure \ref{fig:profiles}C, see inset for schematic), we find a similar shape that also collapses onto the master curve.
Thus, overall crack shape appears to be mostly set by the hydrogel just shrinking to a thickness that depends on the local temperature.
The dashed curve in the Figure is an empirical fit characterizing the collapsed data: $\frac{h}{h_0}=0.18 + 0.24e^{2.25T}+0.53e^{16.12T}$, where $T$ is in Celsius.

\begin{figure}[ht]
    \centering
    \includegraphics[width=1\linewidth]{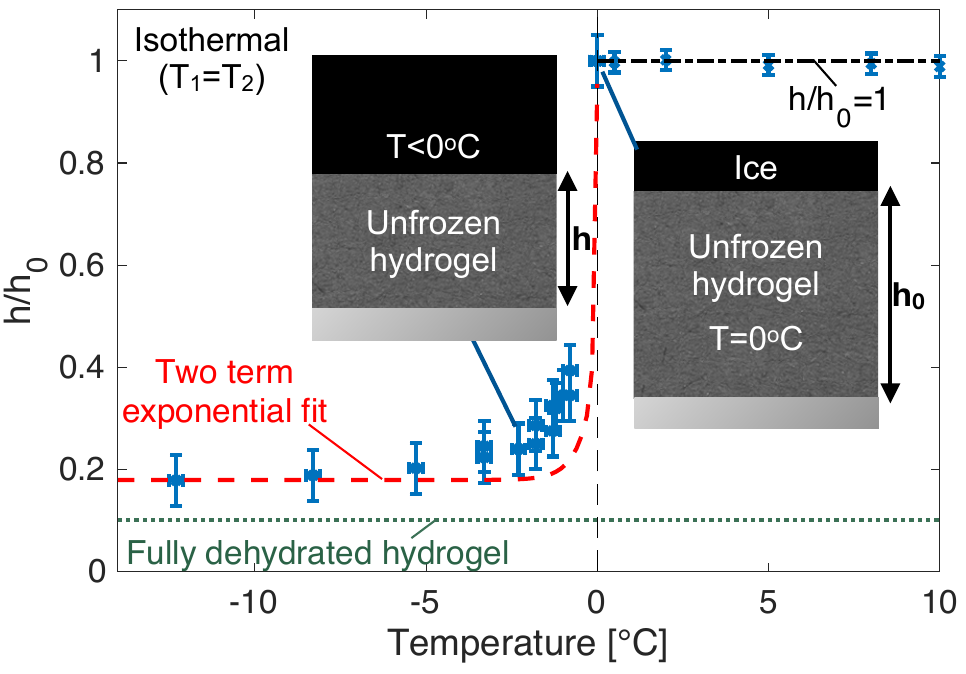}
    \caption{Isothermal shrinkage of freezing hydrogels caused by cryosuction. A layer of PEGDA hydrogel shrinks with increasing undercooling. The gel is in contact with, and equilibrated with ice, and adhered on its bottom side to a rigid glass slide. The red, dashed curve show the empirical fit to the data in Figure \ref{fig:profiles}E.}
    \label{fig:uniaxial}
\end{figure}

The fact that crack shapes largely depend on local temperature suggests that they are dominated by bulk-thermodynamic behavior.
To explore this, we measure the equilibrium between ice and layers of hydrogel with a free, upper surface under isothermal conditions (see schematic in Figure \ref{fig:uniaxial}).
Cryosuction sucks water out of the gel, until the suction is balanced by the dehydrated gel's osmotic pressure.
We characterize this equilibrium by tracking fluorescent particles embedded in the gel layer \cite{kim2021measuring} to  measure how gel thickness, $h(T)$, changes with temperature (Figure \ref{fig:uniaxial}).
At the bulk melting temperature, $T=0^\circ$C, $h=h_0$.
Upon cooling, $h(T)$ reduces monotonically with increasing undercooling, with the hydrogel rapidly losing more than 50\% of its volume in the first degree of undercooling.
Upon further cooling, the dehydration becomes more gradual, as the gel thickness approaches the fully dehydrated limit where only polymer remains (dotted line in Figure \ref{fig:uniaxial}B, calculated using the thickness of the as-prepared hydrogel).

The isothermal shrinkage of the hydrogel with undercooling largely explains the shapes of the delamination profiles in Figure \ref{fig:profiles}.
We add the empirical fit to the data from Figure \ref{fig:profiles}E to Figure \ref{fig:uniaxial} as the red, dashed curve.
These are rather similar, especially at low temperatures.
A perfect match would indicate that fracturing hydrogels in temperature gradients just shrink uniaxially (vertically) until the local osmotic pressure of the gel balances the suction from the overlying ice.
This is true at colder temperatures, where $h$ varies little, so we expect simple uniform shrinkage of the hydrogel.
However, there is less agreement at warmer temperatures.
Here, $h$ changes rapidly, implying that there are additional large shear strains in the delaminating samples (see later for confirmation).
Thus, it is not surprising to see discrepancies.
However, overall, temperature-dependent shrinkage appears to explain the main features of freezing cracks.

\begin{figure}
\centering
    \includegraphics[width=1\linewidth]{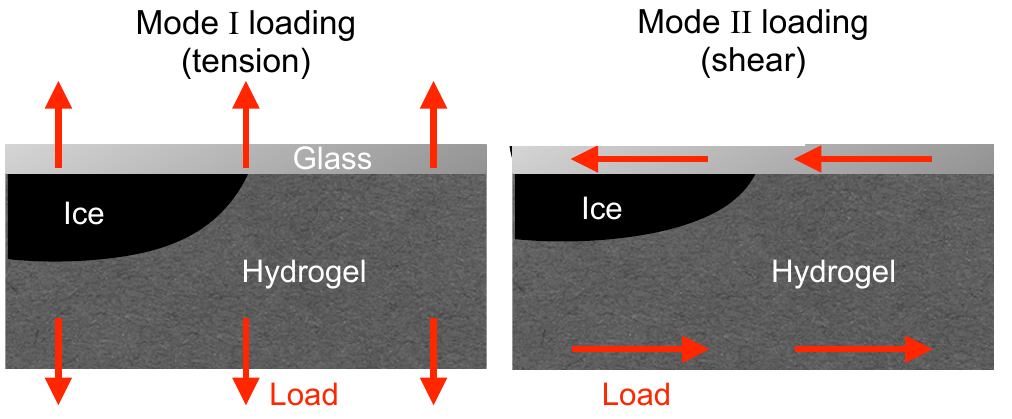}
    \begin{ruledtabular}
\begin{tabular}{l|llll}
   & $K_I$ & $K_{II}$  & $\Gamma_{f}$ & $\Gamma_{p}$\\
   &  [Pa$\mathrm{m}^{1/2}$] &  [Pa$\mathrm{m}^{1/2}$] & [J/$\mathrm{m}^2$] & [J/$\mathrm{m}^2$] \\
   \hline \hline
  PEG silane & $-19\pm5$ & $-55\pm1$ & 0.033 & 1.2 $\pm$ 0.1 \\
  Hexenyl silane & $29\pm6$ & $-137\pm3$ &0.19 & 1.4 $\pm$ 0.3\\
  Methacrylated silane & $150\pm14$ & $-305\pm 17$ & 1.1 & 2.6 $\pm$ 0.34\\
\end{tabular}
\end{ruledtabular}
\caption{\label{tab:results} Top: schematics showing the 2d modes of fracture. The stress intensity factors $K_I$ and $K_{II}$ give the relative magnitudes of the mode I and mode II contributions to fracture. Fitted values are given in the table for the three different silanes used, along with fracture energies from freezing ($\Gamma_{f}$) and bulk-peeling tests ($\Gamma_{p}$). A negative $K_I$ is surprising, as it indicates compression at the crack tip, but this is known to be possible in interfacial cracks \cite{comninou1990overview}.} 
\end{figure}

\begin{figure}
    \centering
    \includegraphics[width=1\linewidth]{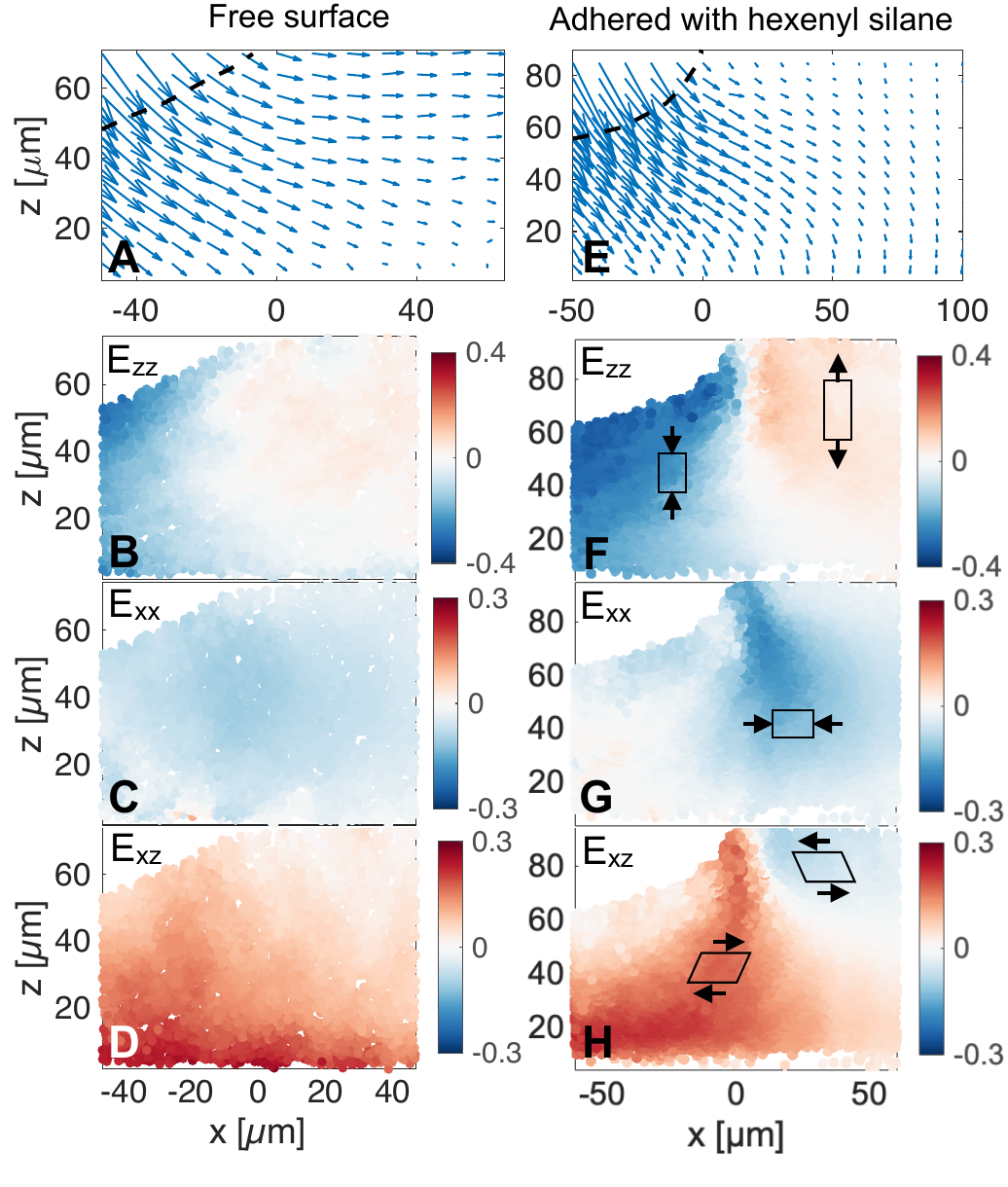}
    \caption{Displacement and strain fields in freezing hydrogel samples. (A-D) A hydrogel with a free upper surface. (A) shows the displacement field near the crack tip (arrows are to scale with the figure axes) while (C-D) show the corresponding strain fields, $E_{xx},E_{zz},E_{xz}$. 
    (E-H) The same data but for freezing of a hydrogel adhered to hexenyl-silane treated glass.}
    \label{fig:strains}
\end{figure}

To get insight into how ice actually breaks the hydrogel, we look closer to the crack tip.
Here, displacements and strains are controlled by interfacial fracture properties, and reveal the specific form of the loading that drives fracture.
Indeed, they distinguish between Mode I loadings acting to push crack faces apart, Mode II loadings acting to shear crack faces past each other, and mixed mode loadings that combine the two behaviors (e.g. Figure \ref{tab:results}) \cite{hutchinson1991mixed}.
To determine hydrogel strains, we track the displacements of embedded nanoparticles in the hydrogel between the initial, stress-free state and the frozen, deformed state, using a large-strain tracking algorithm \cite{kim2021measuring}.
We calculate the local deformation gradient, $\mathbf{F}$, with a linear fit to the local displacement field.
Then, the Green-Lagrange finite strain tensor is $\mathbf{E}=(\mathbf{F}^T\mathbf{F}-\mathbf{I})/2$, where $\mathbf{I}$ is the identity matrix (see Supplement for more details).
Examples of the resulting (grid-interpolated) displacement fields, and the corresponding strains $E_{zz}$, $E_{xx}$ and $E_{xz}$ are shown in Figure \ref{fig:strains}A,E, for a sample with a free upper surface (see schematic in Figure \ref{fig:profiles}C), and the hexenyl-silane sample from Figure \ref{fig:profiles}C.

The strain fields reveal a surprisingly strong shear peeling the hydrogel off the overlying glass.
Figure \ref{fig:strains} shows the strain fields for the free-surface sample (B-D), and the adhered sample (F-H).
Differences between these two data sets can be attributed to the fracture process that only occurs in the adhered sample.
In the free-surface sample, Figure \ref{fig:strains}B shows how the hydrogel contracts vertically under the ice ($x<0$), while only slightly expanding under the water ($x>0$).
There is also a rather uniform horizontal contraction, likely driven by dehydration of the hydrogel (Figure \ref{fig:strains}C), and a positive shear strain that is concentrated at the constrained bottom of the hydrogel layer (Figure \ref{fig:strains}D).
In the adhered sample, there is again vertical contraction of the hydrogel under the ice, this time accompanied by a significant vertical stretching directly ahead of the crack tip (Figure \ref{fig:strains}F).
We also see horizontal contraction, qualitatively similar to what we see in the free-surface sample, but more concentrated near the crack tip (Figure \ref{fig:strains}G).
The most significant difference is the appearance of a strong negative shear strain ahead of the crack tip -- a signature of a `peeling' Mode II crack loading (Figure \ref{fig:strains}H).
This shear can also be seen in the displacement field in Figure \ref{fig:strains}E.
In short, adhesion concentrates stresses and creates both vertical tension and shear at the crack tip.

We determine whether tension or shear drives fracture by measuring the magnitude of Mode I and II loadings at the crack tip.
We do this by fitting the displacement field around crack tips to the asymptotic field predicted by Linear Elastic Fracture Mechanics (LEFM) for interface cracks.
We perform the fitting in a $30\times 20\mu$m box directly to the right of the crack tip.
This box is chosen so that it is small relative to the thickness of the layer, and is positioned where the strains in the free-surface sample are small.
As the fitted strains are not too large here, we anticipate that LEFM locally applies.
The asymptotic, fitted field for interface cracks is more complex than the field for cracks in homogeneous materials, but we can still extract effective `stress intensity factors of classical type', $K_I$ and $K_{II}$, that characterize fracture, just like standard stress intensity factors \cite{rice_elastic_fracture_1988}.
These represent the contribution of Mode I and Mode II loadings, respectively, at the scale at which the crack is loaded -- here, $O(10\mu\mathrm{m})$ (see the Supplement for further details).
The results are given in the table in Figure \ref{tab:results}.
We see that $|K_{II}|\gg|K_I|$.
\emph{i.e.} shear, not wedging open by ice, drives crack growth.

Our results also yield measurements of the fracture energy, $\Gamma_{f}$, of the hydrogel-glass interface during freezing.
$\Gamma_{f}=(K_I^2+K_{II}^2)/[2\mu(1+\nu)\cosh \pi \epsilon]$, where $\nu$ is the drained Poisson ratio of the gel and $\epsilon=-\log(3-4\nu)/(2\pi)$ \cite{hutchinson1991mixed}.
The calculated values (Figure \ref{tab:results}) match our expectation that PEG silane would have much weaker adhesion than the two coupling-agent silanes.
Furthermore, the adhesion energy to the PEG-silanized glass is equal to the interfacial energy of an ice/water interface (0.033$\mathrm{J}/\mathrm{m}^2$ \cite{ketcham_experimental_1969}),
which we expect to be very similar to the ice/hydrogel interfacial energy.
This suggests that the fracture energy in this case is just the interfacial energy required to create new ice/hydrogel interface upon delamination.
Interestingly, all the measured adhesion energies are somewhat smaller than fracture energies, $\Gamma_{p}$, measured by peeling tests on bulk, notched samples at room temperature (see Figure \ref{tab:results}, Materials \& Methods).
One potential reason could be the different temperatures in the two tests.
Another is that the peeling experiments are, necessarily, significantly faster than the quasi-static freezing experiments.
This is important as gel fracture is known to often be rate-sensitive \cite{naassaoui2018poroelastic}.
Finally, in the peeling tests there are sources of dissipation due to processes like visco- and poro-elasticity in the bulk sample, friction, and slipping at the grips -- which can all increase the measured fracture energy.
By contrast, the freezing experiment directly probes only the fracture energy. Thus, $\Gamma_f$ may be closer to the true thermodynamic fracture energy than $\Gamma_p$.

\begin{figure}[htbp]
    \centering
    \includegraphics[width=1\linewidth]{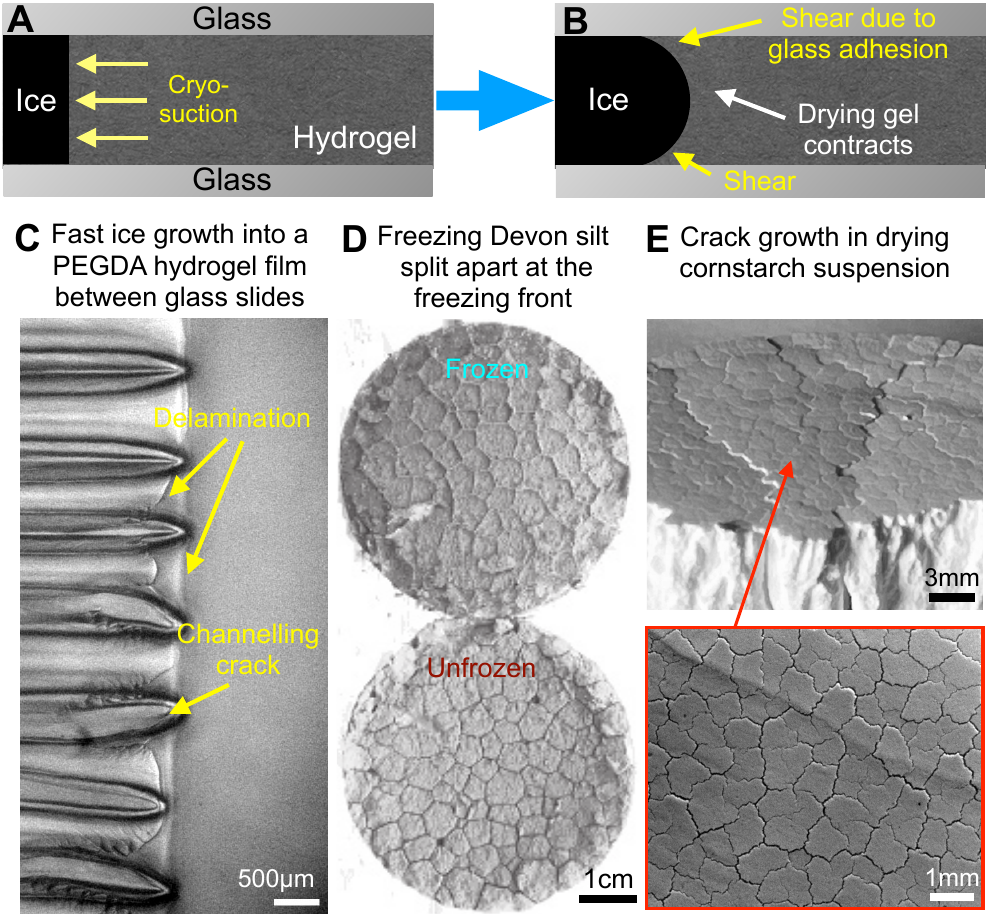}
    \caption{The freezing/drying analogy.  A,B) Cryosuction drives suction of water out of a hydrogel towards cold ice. This dehydrates the hydrogel, leading to shear generation at the glass/hydrogel interface, which drives fracture. C) When ice is grown quickly into the same hydrogel-filled cell used for experiments above, we see both delamination and bulk, channeling cracks. D) A frozen plug of Devon silt, frozen from the top down, and split open at the freezing front (from \cite{arenson2008new}, with permission). The two images are the frozen and unfrozen halves of the sample. E) A drying cornstarch suspension cracks in a similar pattern to the freezing silt. The suspension (50wt\% Xinliang cornstarch in water,  thickness: 10mm) was dried on a hot plate at 50$^\circ$C for 24 hours.} 
    \label{fig:mechanism}
\end{figure}

The underlying mechanism for fracture is gel dehydration caused by cryosuction.
A potential reason for the strong mode II component to fracture would be the ice directly exerting shear on the hydrogel. 
In this case, we would expect to see large shear strains directly under the ice crack in Figure \ref{fig:strains}H.
However, the shear there is rather small, ruling out this mechanism.
Instead, we propose that shear arises due to cryosuction locally dehydrating the hydrogel around the ice.
To the left of the crack tip, this dehydration can be accommodated by the hydrogel simply shrinking in the $z-$direction.
To the right of the crack tip, adhesion prevents shrinkage in the $z-$direction, and the gel contracts laterally, yielding the rightward displacements in Figure \ref{fig:strains}A.
This resulting shear drives the Mode II fracture.
The mechanism is shown schematically for a simpler geometry in Figure \ref{fig:mechanism}(A,B).

Strong evidence for the freeze-fracture as a dehydration process comes from comparing the behaviors of freezing-induced and drying induced-fracture.
For example, drying our sample cells from one end at room temperature induces very similar delamination and strain fields to freezing (see the Supplement).
Furthermore, just like drying, freezing can also drive crack growth into the bulk of a material.
Figure \ref{fig:mechanism}C shows the result of fast ice growth into a 150$\mu$m-thick, hydrogel-filled sample cell.
Here, the ice is advancing into the hydrogel at 10$\mu$m/s, in a temperature gradient $g_T=13^\circ$C/mm.
At these high rates of freezing, we see evenly-spaced cracks that channel through the bulk of the hydrogel (see also Supplementary video 2).
These are very similar to channeling cracks that have been observed in drying films (e.g. \cite{routh2013drying,dufresne2003flow}).
Based on these similarities, we anticipate that bulk freezing cracks similarly grow by dehydrating the nearby material, causing differential stresses that drive propagation \cite{style_ice-lens_2011,routh2013drying}.
In fact, the more one looks into the structure of freezing cracks, the more one finds evidence of a strong freezing/drying analogy.
For example, Figure \ref{fig:mechanism}D,E shows strong visual similarities between the crack patterns in freezing colloidal suspensions (silt, and drying cornstarch respectively).
We anticipate that this topic of bulk freezing fracture is a rich topic for future research.

In conclusion, imaging ice growth into model hydrogels shows that it is not ice expansion upon freezing that drives damage.
Instead, ice growth locally dehydrates the hydrogel, leading to drying-induced stresses that ultimately cause damage.
This mechanism is very similar to that underlying desiccation fracture in brittle, drying materials.
The actual microscopic fracture process is governed by classical fracture mechanics.
Macroscopically, the shape of the cracks is set by a simple balance between cryosuction and gel elasticity (Figure \ref{fig:profiles}).

Our results have important implications for understanding freezing.
For example, we can apply insights from desiccation fracture to understand the freezing process:
In desiccation, there is a strong rate-dependence of fracture, with faster drying leading to large suctions and differential drying -- both of which cause fracture-inducing stresses \cite{style2011mud,goehring2009nonequilibrium,routh2013drying}. 
Our knowledge of this process should shed new light on freezing damage, which is similarly known to be highly rate-dependent \cite{cabrera2020cryopreservation,engelmann_plant_2004}.
As another example, drying films bonded to stiffer substrates are much more likely to break than free-standing films, as they develop residual stresses as they dry out \cite{routh2013drying}.
By analogy, we expect that freezing damage will depend strongly upon how the freezing sample is constrained or confined.
In the opposite direction, the freezing/drying analogy may also help with investigating complex drying problems.
It is hard to perform precise desiccation experiments due to difficulties in spatially controlling humidities and drying rates \cite{goehring2015desiccation}.
Freezing offers an alternative approach to obtain precise control of drying experiments, simply by controlling temperature.

In future, it will be important to understand the role of factors such as solutes and material properties upon the freezing process.
Here, we have avoided additional solutes, to simplify our model system.
However, these are generally present, and can strongly affect freezing behavior \cite{schollick_segregated_2016}.
In particular, ice-active chemicals like ice-binding proteins, ice-nucleating agents or cryopreservants will likely have dramatic effects on freezing damage \cite{deller2014synthetic,fitzner2015many,tas_freezer_2021,bar_dolev_ice-binding_2016,kanji_overview_2017}. 
Furthermore, gel properties could also affect freezing.
For example, physical hydrogels and many biological materials can creep, reducing stress concentrations, \cite{long_fracture_2016,baumberger2006fracture}, while tough hydrogels will be much less brittle than the gels we use here \cite{haque2012super,long_fracture_2016}.
Polymer content should affect fracture, as this determines the degree of drying that a gel can undergo.
Finally, hydrogel mesh size should also play an important role, as this determines when ice can penetrate into the bulk of a hydrogel.

\section{Materials and Methods}
The silicone gel is made following the recipe of \cite{kim_extreme_2020}, and comprises a mixture of vinyl-terminated, silicone polymer chains (DMS-V31, Gelest) cross-linked with a methylhydrosiloxane–dimethylsiloxane copolymer (HMS-301, Gelest). Polymerization is catalysed using Karstedt’s catalyst  (SIP6831.2,  Gelest), and samples are cured at $60^\circ$C for several days to ensure complete reaction.
The PEGDA hydrogel is fabricated by making a $10\mathrm{vol}\%$ polyethylene glycol diacrylate (molecular weight 700 Da, Sigma-Aldrich) solution in water.
Then we add 2-hydroxy-2-methylpropiophenone UV-initiator as $0.0005\mathrm{vol}\%$ of the solution, and fluorescent nanoparticles (200nm red, carboxylate-modified Fluospheres, Thermo Fisher Scientific) as $0.001\mathrm{vol}\%$ of the solution.
Finally we crosslink the gel in the glass cell under 365nm UV light for 1 hour.
We mechanically characterize the gels by indenting bulk samples using a TA.XT Plus texture analyzer with a 500g load cell (Stable Microsystems).
The silicone gel shear modulus ($\mu=30$kPa) is measured with a 1-mm radius, cylindrical indenter.
The hydrogel shear modulus ($\mu=37$kPa) and drained Poisson ratio ($\nu=0.17$) are measured using a 1.6-mm radius, spherical indenter, following \cite{hu_using_2010}. 
The experiments are performed on a hydrogel submerged in water in an ice bath, to obtain properties of a gel at 0$^\circ$C.

The cell is made of two glass microscope slides, bonded together with spacers.
To control the attachment of the hydrogel to the inside of the cell, we silanize the microscope slides before cell assembly.
We clean the slides with ethanol and water, dry them, and then activate them in a UV ozone cleaner for 10 minutes (ProCleaner Plus, Bioforce Nanosciences).
We mix together 900$\mu$l ethanol, 50$\mu$l de-ionized water, 50$\mu$l of glacial acetic acid and finally add $3\mu$l of the appropriate silane.
After 5 minutes of resting time, we apply 30$\mu$l of the mixture to each glass slide, and leave them for 3 minutes to allow full surface coverage.
Next we quench the reaction immersing the slides in ethanol, before drying the slides on a hot plate at 110$^\circ$C for 10 minutes to complete the silanization reaction.
Glass slides were stored in a dry box (humidity below 20\%) until use.

The samples are imaged with a Nikon Ti2 Eclipse microscope with a 20x air objective and a 3i Spinning Disk Confocal system.
We use a 561nm laser to image fluorescent tracer particles, and brightfield imaging for observing ice/water interface positions (e.g. Figure \ref{fig:schem}).

We performed bulk peeling tests by peeling hydrogel films off glass surfaces silanized with different silanes.
By peeling at a fixed angle ($\theta=10^\circ$), we calculate the fracture energy as $ \Gamma = (F_{c}/w)( 1-\cos\theta)$.
Here, $F_c$ is the steady state peeling force and $w=18$mm is the width of the hydrogel contact line \cite{bartlett2023peel}.
The hydrogel was prepared with a thickness of $0.5$mm on the glass surface, and its upper surface was attached to a backing layer of 3M Scotch tape using a minimal amount of superglue (Deli 502).
Peel tests were performed on a dynamic mechanical analyzer (DMA850, TA Instruments).
All tests were done on as-prepared hydrogels at room temperature. 

\section{Acknowledgements}
We acknowledge support from an ETH Research Grant (ETH-38~18-2), and the Swiss National Science Foundation (200021-212066), and helpful conversations with Nan Xue, Stefanie Heyden and Rong Long. 
SY would like to acknowledge financial support from China Scholarship Council (No. 202106020115). This work is also partially supported by the National Natural Science Foundation of China (Nos. 12072010 and 11674019) and the Fundamental Research Funds for the Central Universities (YWF-22-K-101).


\end{document}